\documentclass[12pt]{article}
\usepackage[dvips]{color}
\usepackage{epsfig}
\usepackage{amsmath}
\usepackage{cite}
\usepackage{color}
\usepackage{subfigure}

\begin{document}
\begin{center}
\Large{\bf Pleasant behavior of swampland conjectures in the face of specific inflationary models}\\
\small \vspace{1cm}{\bf S. Noori Gashti$^{\star}$\footnote {Email:~~~saeed.noorigashti@stu.umz.ac.ir}}, \quad
 {\bf J. Sadeghi$^{\star,\dagger}$\footnote {Email:~~~pouriya@ipm.ir}}, \quad {\bf B. Pourhassan$^{\star\star,\dagger}$\footnote {Email:~~~b.pourhassan@du.ac.ir}}, \quad
\\
\vspace{0.5cm}$^{\star}${Department of Physics, Faculty of Basic
Sciences,\\
University of Mazandaran
P. O. Box 47416-95447, Babolsar, Iran}\\
\small \vspace{0.2cm} $^{\star\star}${School of Physics, Damghan University, Damghan, 3671641167, Iran}\\
\small \vspace{0.2cm} $^{\dagger}${Canadian Quantum Research Center 204-3002 32 Avenue Vernon\\ British Columbia V1T 2L7 Canada}\\

\end{center}
\begin{abstract}
Recently researchers have been studying various conditions as swampland criteria in cosmological implications. They have studied many inflation models with different swampland conditions. Occasionally these conjectures are modified and lead to fascinating points. The swampland conjecture is in contradiction with the slow-roll single-field inflation model. So in this paper, we will briefly describe resolving this contradiction with the special method; we consider a small parameter $\lambda$ with respect to a coupling function in inflation and resolve this important antithesis. Then we will discuss a new point of view in studying inflation models according to the swampland criteria. Therefore we introduce some inflation models and challenge the swampland criteria. We specify the allowable range of cosmological parameters as the tensor-to-scalar ratio ($r$), the scalar spectral index ($n_{s}$) according to the latest observable data. Next, we study the new constraints to determine the compatibility or incompatibility of the corresponding model with the swampland conjectures. Finally, we express the results and compare them with the latest observational data by plotting some figures.
\end{abstract}
\newpage
\tableofcontents
\section{Introduction}

Due to problems in standard cosmology such as horizons,  monopole, and flatness, various solutions have been proposed to address these problems, including inflation. With science development, new modifications are accompanied for solutions,
and also, there will exist new questions for researchers. Recent studies in cosmology and other sciences such as particle physics have led to remarkable developments. Cosmic microwave background (CMB) measurements that show fluctuations in matter and energy are always unstable on a large scale. The main cause of these fluctuations is still unknown. Researchers have provided explanations for the reasons for these fluctuations, such as the inflationary universe like all cosmological questions and problems. So, cosmic inflation has been proposed as an amazing pattern for these problems. To solve these problems, the universe has gone through an early phase of accelerated expansion \cite{1,2,3,A,B,C,D,4,5,6,7,8,9,10,11,12,13,14}. Different types of these inflation models with various conditions and their implications in cosmology have been studied thoroughly. The simplest model is single-field inflation with the slow-roll condition. It can be shown that inflation is practical for more than one field. However, the study of single-field inflation models and their consequences is of more interest to researchers \cite{15,16,17,18,19,20}. Recently, the concept of the swampland program has been introduced, which has some conditions as swampland dS conjecture,  distance conjecture,  TCC, etc. One of them is weak gravity conjecture that states gravity is the weakest force in theories coupled with gravity.
On the other hand, a set of theories are placed in the landscape (an area compatible with quantum gravity), and other effective theories are usually placed in an area called the swampland, which is incompatible with quantum gravity. The swampland is wider than the landscape, and it surrounds the landscape. So most effective theories are located in the swampland. To adapt to quantum gravity at low energy levels, they must satisfy the swampland conditions \cite{21,E,F,22,G,H,I,J,23,K,L,M,24,N,O,P,Q,R,S,25,T,26,V,W,X,Y,Z,27,a,b,c,28,29,30,31,32,aa,bb}. Single-field inflation with the slow-roll conditions is usually based on effective low-energy theories derived from Einstein equations. So to be consistent with quantum gravity theories, they must meet several criteria known as swampland criteria.  The swampland distance conjecture, which normalizes the scalar field excursion with reduced Planck mass, and the field space is limited from above. Also the refined swampland dS conjecture, which provides a limit to the potential \cite{26,V,W,X,Y,Z,33,34,cc,35,36}.
Single-field inflation models with slow-roll and swampland conjectures have a small problem, i.e., there is a contradiction between single-field slow-roll inflation with one of the swampland criteria, which we will explain in the following. Researchers have developed a powerful method for this problem by connecting a coupling as Gauss-Bonnet to a potential that solves the problems associated with swampland by a special mechanism using a parameter $(1-\lambda)^{2}$ \cite{37}. Gauss-Bonnet is an exciting theory and also may even solve problems of the universe's singularity. The speed of gravitational waves in such a model differs from the speed of light. In the last decade, different Gauss-Bonnet structures, and their implications have been studied in various cosmological topics \cite{38,d,e,f,39,i,j,40,k,l,m,n,o,41,p,q,42,r,s,43,t,u,44,v,w,45,x,y}. However, regarding the swampland conjecture, an important question is relation of Gauss-Bonnet term with the string theory.
Indeed, Gauss-Bonnet term may reproduced from the first order $\alpha^{\prime}$ correction of the string theory \cite{GBS1, GBS2}. Then, the black hole solutions in the superstring effective field theory with the Gauss-Bonnet term may considered \cite{GBS3}. Also charged black hole solutions in Einstein-Gauss-Bonnet theory with the dilaton field as the low-energy effective theory of the heterotic string has been studied \cite{GBS4}.\\
Important cosmology parameters in inflation are calculated as the scalar spectral index $n_{s}$, and the tensor-to-scalar ratio ($r$) under the slow-roll conditions regarding $\epsilon\ll1$, and $\eta\ll 1$.
With the Gauss-Bennet connection, the slow-roll condition is converted to $(1-\lambda)(\frac{V'}{V})^{2}\ll 1$. So the swampland criteria are satisfied as long as $(1-\lambda)\ll 1$ and compatible with the slow-roll condition. The related results are also well-satisfying. Also in various works in literature determine that the modified theory resolved this like problem. So in this paper, we examine some inflation models from this perspective, and we will show this compatibility between the slow-roll condition and swampland conjecture. Recently in the literature to solve this antithesis, some solutions have been proposed that are related to the tensor to scalar ratio and properties of ($\Delta \phi $), you can see the Refs. \cite{46,47,48,49,50,51,52,53,54,55,56,57,58,59,60,61,62,63,64,65,66,67,68,69,70,71,72,73}.
So this paper is organized as follows.\\
In section 2, we will briefly describe the base equation about inflation and some points related to refined swampland dS conjectures completely. In section 3, we study a new perspective in studying inflation models according to the swampland criteria. We introduce some inflation models, and we investigate the important cosmological parameters as $N$, $r$, and $n_{s}$ with respect to the concept mentioned in this paper. We specify the allowable area of cosmological parameters such as ($r$), ($n_{s}$), and reconstructed new restrictions as ($C_{1}-r$), ($C_{1}-n_{s}$), ($C_{2}-r$), and ($C_{2}-n_{s}$).
We determine the allowable range for the swampland component corresponding to the models by plotting some figures and advantaging with the latest observational data. Finally, we state the result in section 4.

\section{The basic equations}

In this section, we study the basic equation according to the concepts presented. Hence, we generally express the action in the following form \cite{74,75,76,77,78,79,80,z,81},

\begin{equation}\label{1}
S=\frac{1}{2}\int\sqrt{-g}d^{4}x\bigg(R-g^{\mu\nu}\partial_{\mu}\phi\partial_{\nu}\phi-2V(\phi)-\xi(\phi)R^{2}_{GB}\bigg),
\end{equation}
where $\xi(\phi)$, $R^{2}_{GB}$ are Gauss-Bonnet coupling function and Gauss-Bonnet term, respectively, which are as follows,
\begin{equation}\label{2}
R^{2}_{GB}=R_{\mu\nu\rho\sigma}R^{\mu\nu\rho\sigma}-4R_{\mu\nu}R^{\mu\nu}+ R^{2},
\end{equation}
and the Gauss-Bonnet coupling function \cite{37},
\begin{equation}\label{3}
\xi(\phi)=\frac{3\lambda}{4V(\phi)+\Lambda_{0}},
\end{equation}
where $\Lambda_{0}\ll(10^{16}Gev)^{4}$, so in this paper we ignore the $\Lambda_{0}$. Also, the observational constraints \cite{3} suggests $0<\lambda<1$ \cite{37}. Indeed, the marginalized $1\sigma-3\sigma$ confidence level contours for scalar spectral index and tensor-to-scalar ratio from Planck 2015
and BICEP2/Keck data \cite{3} suggest the constant $\lambda$ of order unity to match mentioned observations.\\
The slow-roll parameter satisfy the conditions as $\epsilon_{1}=-\frac{\dot{H}}{H^{2}}\ll1$ and $\epsilon_{2}=\frac{\dot{\epsilon_{1}}}{H\epsilon_{1}}\ll1$.
Also, the number of e-folds ($N$) at the horizon, before the end of inflation and according to the potential, is expressed in the following form,
\begin{equation}\label{4}
N\approx\int_{\phi_{e}}^{\phi}\frac{3V}{3V,\phi+4\xi,\phi V^{2}}d\phi=\int_{\phi_{e}}^{\phi}\frac{d\phi}{Q},
\end{equation}
where $Q=\frac{V,\phi}{V}+\frac{4\xi,\phi V}{3}$. According to the equation (\ref{4}), we will have \cite{37}
\begin{equation}\label{5}
N\approx\int_{\phi_{e}}^{\phi}\frac{1}{\sqrt{2\epsilon_{1}}}d\phi.
\end{equation}
The scalar spectral index expressed in the following form \cite{37},
\begin{equation}\label{6}
n_{s}=1-2\epsilon_{1}-\epsilon_{2}
\end{equation}
where the $n_{s}$ is the scalar spectral tilt. Also, the tensor-to-scalar ratio is as following,
\begin{equation}\label{7}
r=16(1-\lambda)\epsilon_{1}.
\end{equation}
The above slow-roll parameters can state with the potential term, which is introduced as follows \cite{82,83,84},
\begin{equation}\label{8}
\epsilon_{1}=\frac{1-\lambda}{2}\left(\frac{V'}{V}\right)^{2},
\end{equation}
and the second parameter is as follows,
\begin{equation}\label{9}
\epsilon_{2}=-2(1-\lambda)\left[\frac{V''}{V}-(\frac{V'}{V})^{2}\right].
\end{equation}

From the two equations (\ref{8}) and (\ref{9}) as well as the matching of potential and swampland conditions, with respect to ($ \lambda $), that be very close to 1, the slow-roll conditions as $ \epsilon_ {1}$ and $ \epsilon_ {2}$ are satisfied. Therefore, it can be concluded that equations (\ref{6}) and (\ref{7}) are correct and valid. Assuming the number of e-folds is equal to 60. So, from equations (\ref{7}) and (\ref{8}), we will have,
\begin{equation}\label{10}
r=8(1-\lambda)^{2}\left(\frac{V'}{V}\right)^{2}.
\end{equation}
From the equation (\ref{10}), the second swampland dS conjecture is satisfied, and the contradiction is resolved. Also, if we consider the $(1-\lambda) $ is very small, the tensor-to-scalar ratio $r$ can be consistent with observable data. The first swampland criteria, according to the Lyth bound, is given by \cite{37},
\begin{equation}\label{11}
\Delta\phi=(1-\lambda)\Delta N \frac{V'}{V}.
\end{equation}
If we ignore the $\lambda$, the second swampland criteria are satisfied, but it is not met for slow-roll single-field inflation. Therefore if the $(1-\lambda)$ be very small, both of the swampland criteria are satisfied. Here with respect to the second swampland conjecture, and $\Delta\phi\leq d$, one can obtain,
\begin{equation}\label{12}
1-\lambda<\frac{d}{c \Delta N}.
\end{equation}
The remarkable point in the concepts presented in this section is the use of the Gauss-Bonnet term. With the help of the Gauss-Bonnet term, it satisfies both swampland criteria as long as the conditions expressed for the $\lambda$ are met. Besides, the tensor-to-scalar ratio ($r$) will be in line with observable data\cite{3}. Therefore, Given all the above concepts, we calculate each of the mentioned cosmological quantities in Einstein's framework for each model separately by introducing several potentials. By calling swampland conjectures in this framework, we will examine the results about each of these variables and the parameter $\lambda$. We compare the results of different models with each other as well as the latest observable data. We also plot some figures to determine the range associated with each of these cosmological parameters.

\section{The Inflation Models}
In this section, we introduce some inflation models as follows, and apply the concepts expressed in the previous section to these inflation models \cite{85,86,87},\\\\
$\bullet U(\psi)=V_{0}\bigg(1-\bigg[\psi/\mu\bigg]^{n}\bigg)$\\
$\bullet U(\psi)=V_{0}\tanh^{2n}\bigg[\psi/\sqrt{6\alpha}\bigg]$\\

We calculate the cosmological parameters as $r$, $n_{s}$, etc., and determine the range of each of these parameters by plotting some figures.  We also use the small parameter $\lambda$ to solve the problems associated with single-field inflation with slow-roll and swampland conditions. Hence, we first introduce the swampland conjectures, then we describe the results associated with any potential. As mentioned above, swampland criteria are expressed in the following form,
\begin{equation}\label{13}
\frac{\Delta\phi}{M_{pl}}<\mathcal{O}(1),
\end{equation}
and
\begin{eqnarray}\label{14}
M_{pl}\frac{|U'|}{U}>C_{1},\nonumber\\
\frac{min (|\nabla_{i}\nabla_{j}U|)}{U}=M_{pl}^{2}\frac{|U''|}{U}\leq -C_{2}.
\end{eqnarray}

\subsection{Case I}
As a first case, we consider a hilltop attractor potential, So we have
\begin{equation}\label{15}
U(\psi)=V_{0}\bigg(1-\bigg[\psi/\mu\bigg]^{n}\bigg).
\end{equation}
With respect to the above equation and swampland conjectures in equation (\ref{14}), one can obtain,
\begin{equation}\label{16}
\frac{U'}{U}=\frac{n(1+\frac{1}{-1+(\frac{\psi}{\mu})^{n}})}{\psi}>C_{1},
\end{equation}
and
\begin{equation}\label{17}
\frac{U''}{U}=\frac{(-1+n)n(\frac{\psi}{\mu})^{n}}{\psi^{2}(-1+(\frac{\psi}{\mu})^{n})}>-C_{2},
\end{equation}
where $U'$, $U''$ are the first and second derivatives of $\psi$. The tensor-to-scalar ratio and scalar spectral tilt for this model in Einstein framework can obtain with respect to the above equations as
\begin{equation}\label{18}
r=\frac{8(1-\lambda)^{2}}{\mu^{2}(\mathcal{W}^{\frac{(1-\lambda)-1}{(1-\lambda)-2}}-\mathcal{W}^{-\frac{1}{(1-\lambda)-2}})^{2}}
\end{equation}
and
\begin{eqnarray}\label{19}
n_{s}&=&1-\frac{2(1-\lambda)-1}{((1-\lambda)-2)N+(1-\lambda)-1}\nonumber\\
&-&\frac{(1-\lambda)\bigg\{(5(1-\lambda)-2)\mathcal{W}^{\frac{1-\lambda}{((1-\lambda)-2)}}
-2((1-\lambda)-1)\bigg\}}{\mu^{2}\mathcal{W}\bigg\{-2\mathcal{W}^{\frac{1-\lambda}{((1-\lambda)-2)}}+\mathcal{W}^{\frac{2(1-\lambda)}{((1-\lambda)-2)}}+1\bigg\}},
\end{eqnarray}
where $\mathcal{W}=\frac{(1-\lambda)[((1-\lambda)-2)N+(1-\lambda)-1]}{\mu^{2}}$. Here we invert equation (\ref{18}) in terms of ($r$) and equation (\ref{19}) in terms of ($n_{s}$) as $N-r$, and $N-n_{s}$. By considering the equations (\ref{4}), and (\ref{16})-(\ref{19}) the new expressions will be obtained in terms of ($r$) and ($n_{s}$). So we will have
\begin{equation}\label{20}
\begin{split}
&C_{1}-n_{s}=1/\mu 2^{1/n}n\bigg((-1+\lambda^{2})(-2\hspace{2mm} _{2}F_{1}\bigg[1,-1/2+1/n,1/2+1/n,4\lambda/-1+\lambda\bigg]\\
&+2\hspace{2mm}_{2}F_{1}\bigg[1,1/n,1+1/n,4\lambda/-1+\lambda\bigg]\\
&-n \hspace{2mm}_{2}F_{1}\bigg[1,1/n,1+1/n,4\lambda/-1+\lambda\bigg]\bigg/(-2+n)n(1-4\lambda)\bigg(\lambda-\lambda^{2}+4^{-\frac{1+\lambda}{-1+\lambda}}\times\\
&\bigg[\frac{-5+n_{s}+5\lambda+2\lambda n_{s}+\sqrt{57-42n_{s}+9n_{s}^{2}-18\lambda-90\lambda n_{s}+36\lambda n_{s}^{2}+25\lambda^{2}-12\lambda^{2}n_{s}+36\lambda^{2}n_{s}^{2}}}{-2+n_{s}+2\lambda n_{s}}\bigg]^{\frac{1+\lambda}{-1+\lambda}}\\
&\times\mu^{2}\bigg)\bigg)
\end{split}
\end{equation}
and
\begin{equation}\label{21}
\begin{split}
&\mathcal{A}=2\hspace{2mm}_{2}F_{1}\bigg[1,-\frac{1}{2}+\frac{1}{n},\frac{1}{2}+\frac{1}{n},1+\frac{1}{-1+4\lambda}\bigg]+(-2+n)\\
&\times \hspace{2mm}_{2}F_{1}\bigg[1,\frac{1}{n},1+\frac{1}{n},1+\frac{1}{-1+4\lambda}\bigg]\\
&\mathcal{B}=(4^{1+\lambda/-1+\lambda}(-1+\lambda)\lambda\\
&-(\frac{-5+n_{s}+5\lambda+2\lambda n_{s}+\sqrt{57-6n_{s}(7+\lambda)(1+2\lambda)+\lambda(-18+25\lambda)+9(n_{s}+2n_{s}\lambda)^{2}}}{-2+n_{s}+2\lambda n_{s}})^{\frac{(1+\lambda)}{(-1+\lambda)}}\mu^{2})\\
&C_{2}-n_{s}=\bigg\{4^{1/n}(-1+n)n\bigg(\frac{2^{2+\frac{4}{-1+\lambda}}(-1+\lambda^{2})\mathcal{A}}{(-2+n)n(-1+4\lambda)\mathcal{B}}\bigg)^{-2/n}\times \bigg(\bigg[\frac{2^{(3+\lambda)/(-1+\lambda)}(-1+\lambda)\mathcal{A}}{(-2+n)n(-1+4\lambda)\mathcal{B}}\bigg]^{1/n}\bigg)^{n}\bigg\}\\
&\bigg/\bigg(\mu^{2}\bigg[-1+\bigg(\bigg[\frac{2^{(3+\lambda)/(-1+\lambda)}(-1+\lambda^{2})\mathcal{A}}{(-2+n)n(-1+4\lambda)\mathcal{B}}\bigg]^{1/n}\bigg)^{n}\bigg]\bigg)
\end{split}
\end{equation}
and swampland conjectures in terms of $r$ as,
\begin{equation}\label{22}
\begin{split}
&\mathcal{X}=2\hspace{2mm}_{2}F_{1}\bigg[1,-\frac{1}{2}+\frac{1}{n}+\frac{1}{2}+\frac{1}{n},1+\frac{1}{-1+4\lambda}\bigg]+(-2+n)\\
&\times \hspace{2mm}_{2}F_{1}\bigg[1,\frac{1}{n}+1+\frac{1}{n},1+\frac{1}{-1+4\lambda}\bigg]\\
&\mathcal{Y}=(-2+n)n(-1+4\lambda)\bigg[\sqrt{r}(-1+\lambda)\lambda+2^{3(1+\lambda)/2\lambda}\sqrt{(-1+\lambda)^{2}}(-\frac{\sqrt{(-1+\lambda)^{2}}}{\mu\sqrt{r}})^{1/\lambda}\mu\bigg]\\
&C_{1}-r=1/\mu 2^{1/n}n\bigg(\frac{\sqrt{r}(-1+\lambda^{2})\mathcal{X}}{\mathcal{Y}}\bigg)^{-1/n}\times\bigg[1+\frac{1}{-1+\bigg(2^{-1/n}\bigg[\frac{\sqrt{r}(-1+\lambda^{2})\mathcal{X}}{\mathcal{Y}}\bigg]^{1/n}\bigg)^{n}}\bigg]
\end{split}
\end{equation}
and
\begin{equation}\label{23}
\begin{split}
&\mathcal{G}=-2\hspace{2mm}_{2}F_{1}\bigg[1,-\frac{1}{2}+\frac{1}{n}+\frac{1}{2}+\frac{1}{n},\frac{4\lambda}{-1+4\lambda}\bigg]\\
&+2\hspace{2mm}_{2}F_{1}\bigg[1,\frac{1}{n}+1+\frac{1}{n},\frac{4\lambda}{-1+4\lambda}\bigg]\\
&-n \hspace{2mm}_{2}F_{1}\bigg[1,\frac{1}{n}+1+\frac{1}{n},\frac{4\lambda}{-1+4\lambda}\bigg]\\
&\mathcal{K}=\bigg((-2+n)n(-1+4\lambda)\bigg[\lambda-\lambda^{2}+2^{3(1+\lambda)/2\lambda}\bigg(-\frac{\sqrt{1-2\lambda+\lambda^{2}}}{\sqrt{r}\mu}\bigg)^{(1+\lambda)/\lambda}\mu^{2}\bigg]\bigg)\\
&C_{2}-r=\bigg(2^{2/n}(1+n)n\bigg(\bigg[(-1+\lambda^{2})\bigg\{\mathcal{G}\bigg\}\bigg]\bigg/\mathcal{K}\bigg)^{-2/n}\bigg(2^{-1/n}\bigg[\bigg((-1+\lambda^{2})\bigg\{\mathcal{G}\bigg\}\bigg)\bigg/\mathcal{K}\bigg]^{1/n}\bigg)^{n}\bigg)\\
&\bigg/\bigg(\mu^{2}\bigg[-1+\bigg(2^{-1/n}\bigg[\bigg((-1+\lambda^{2})\bigg\{\mathcal{G}\bigg\}\bigg)\bigg/\mathcal{K}\bigg]^{1/n}\bigg)^{n}\bigg]\bigg).
\end{split}
\end{equation}

Now, we can determine these cosmological parameters range by plotting some figures.\\

\begin{figure}[h!]
 \begin{center}
 \subfigure[]{
 \includegraphics[height=6cm,width=6cm]{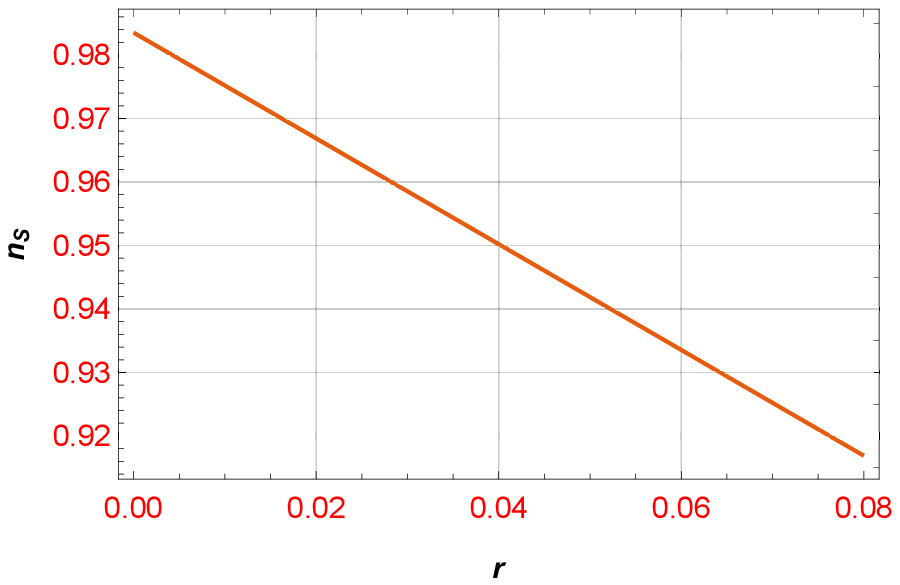}
 \label{1a}}
  \caption{\small{The plot of $n_{s}$ in term of $r$.}}
 \label{1}
 \end{center}
 \end{figure}

 \begin{figure}[h!]
 \begin{center}
 \subfigure[]{
 \includegraphics[height=3.5cm,width=3.5cm]{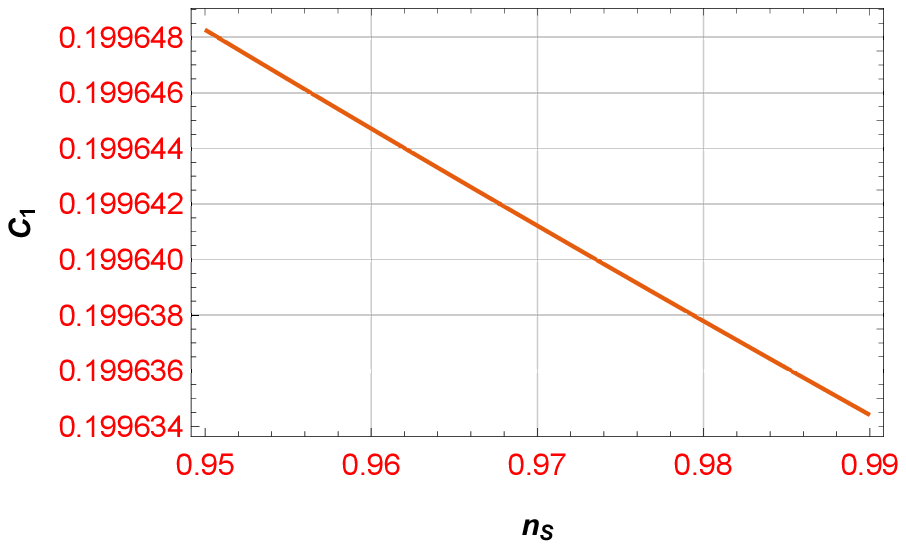}
 \label{2a}}
 \subfigure[]{
 \includegraphics[height=3.5cm,width=3.5cm]{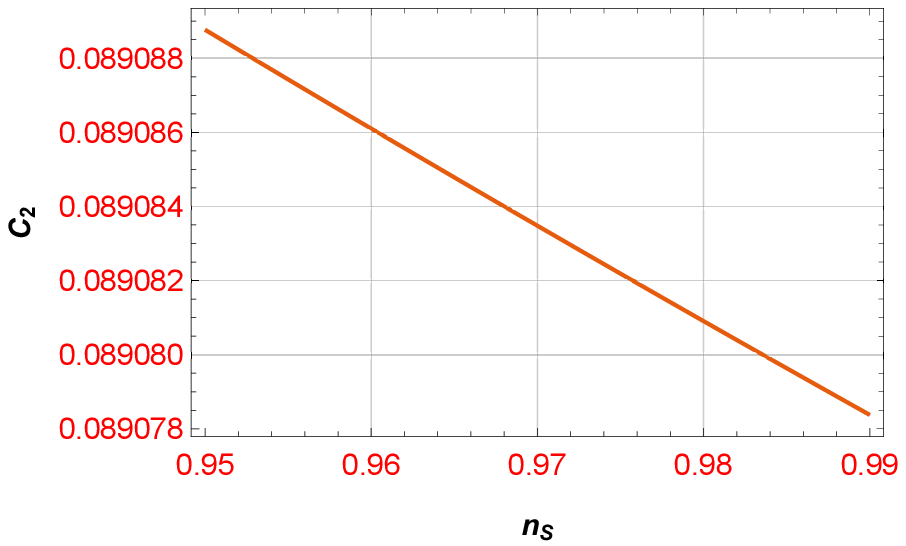}
 \label{2b}}
 \subfigure[]{
 \includegraphics[height=3.5cm,width=3.5cm]{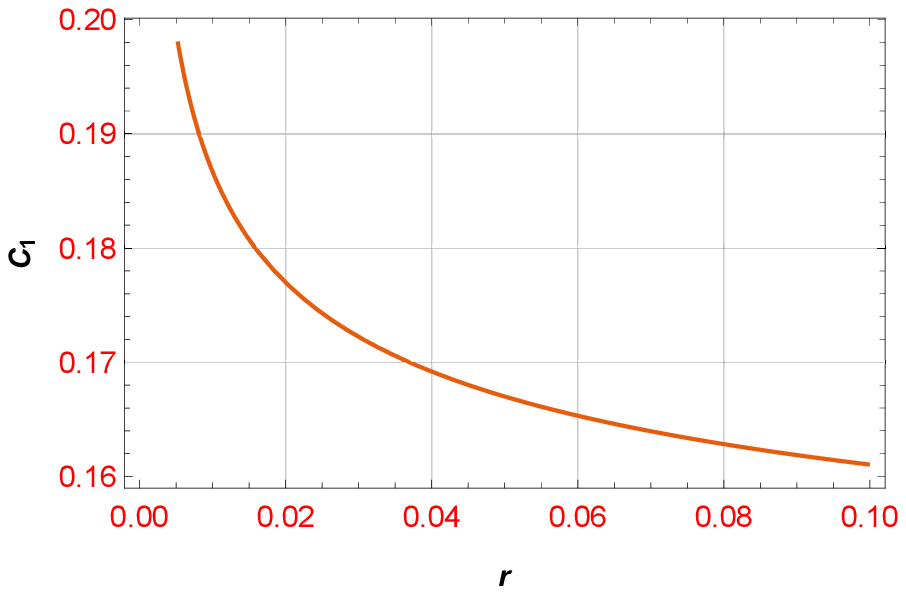}
 \label{2c}}
  \subfigure[]{
 \includegraphics[height=3.5cm,width=3.5cm]{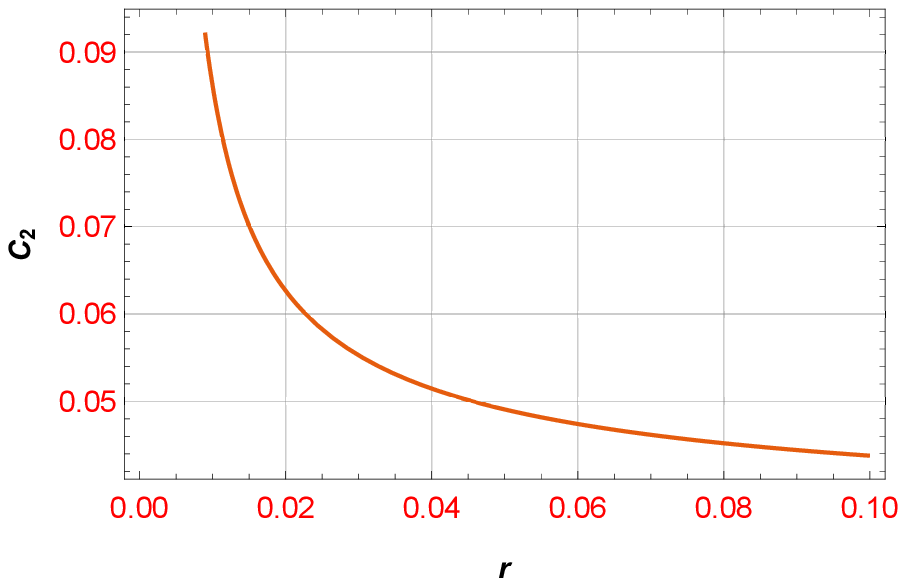}
 \label{2d}}
  \caption{\small{The plot of ($C_{1}$ and $C_{2}$) in term of ($n_{s}$), and (r) in Fig. \ref{2a}, \ref{2b} and Fig. \ref{2c}, \ref{2d}, respectively. }}
 \label{2}
 \end{center}
 \end{figure}

 \begin{figure}[h!]
 \begin{center}
 \subfigure[]{
 \includegraphics[height=6cm,width=6cm]{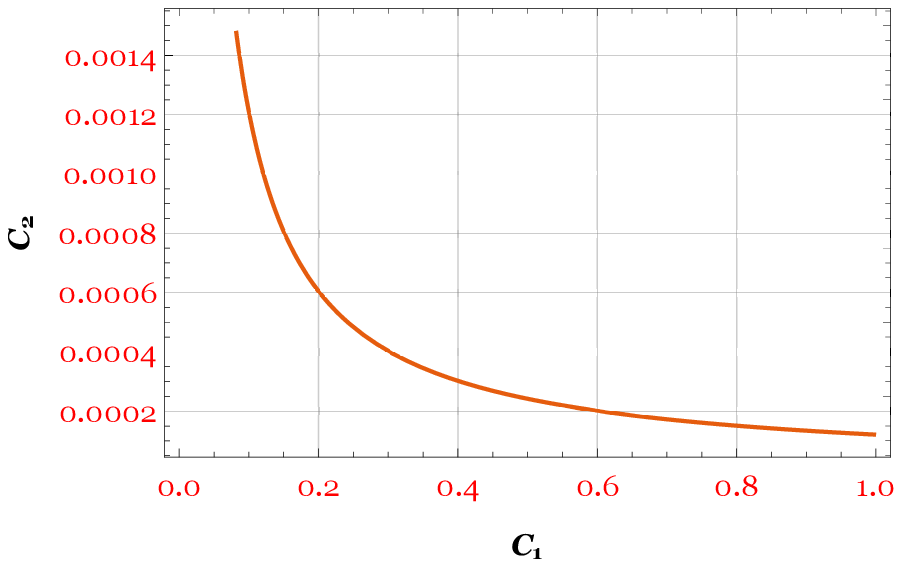}
 \label{3a}}
  \caption{\small{the constraint of ($C_{2}-C_{1}$).}}
 \label{3}
 \end{center}
 \end{figure}

As you can see in the figures of model I, we first plotted the two important cosmological parameters; (tensor-to-scalar ratio) $r$ in terms of $n_{s}$ in figure \ref{1}. As shown in figure \ref{1}, the allowable areas of these two parameters are well determined for this model. But the important point that we should mention is the study of swampland conditions according to important cosmological parameters such as $r$, and $n_{s}$ with advantaging the parameter $ \lambda $. we will assume $ \lambda=0.990$, $\mu<1,\simeq0.33$, $n=3$. Figure \ref{2} shows the changes in the swampland conjectures component,i.e., ($C_{1}$), and ($C_{2}$) in terms of two important cosmological parameters, ($n_{s}$), and ($r$). The changes of ($C_{1}$), and ($C_{2}$) in terms of ($n_{s}$) in figure \ref{2a}, and \ref{2b} are specified.
 The allowable range of these parameters is well determined.
As shown in the figures, the swampland conjecture components are within the allowable range. As stated in the literature, ($C_{1}$) and ($C_{2}$) are positive and unit order.
Also, the values of the $C_{2}$ are smaller than the $C_{1}$. Figures \ref{2c} and \ref{2d} show the variations of two components,  ($C_{1}$), and ($C_{2}$) in terms of $r$. The allowable range associated with each of these components is well represented.
Here, the important point is the role of parameter $\lambda$ to resolve the contradiction between parameter $r$ and swampland conjectures, which has eliminated this defect well.
Using equations (\ref{4}), (\ref{16}), (\ref{17}), and (\ref{19}), we will examine another constraint from the perspective of swampland conjectures and obtain the allowable range or the extent to which the above-inflation model can be valid in swampland conjectures.
We compute the $C_{1}^{2}C_{2}^{2}$ according to the condition $f>C_{1}^{2}C_{2}^{2}$ to investigate this constraint. With respect to the latest observable data and the mentioned condition, the allowable range for swampland conjectures of this model is specified. In figure \ref{3}, we plot the plan ($ C_{2}-C_{1} $).
The allowable area for swampland conjectures regarding the model I for the mentioned constant parameters will be $c<0.000910209$.
In the remainder of this paper, we examine the changes of cosmological parameters with respect to another potential. We determine the range of these parameters.
It should be noted that the exact values of each of these cosmological parameters, such as the scalar spectral index ($ n_{s}$), the tensor to scale ratio ($r$), etc., can be examined.
Our main focus is to study the swampland conjecture behavior for different values and coefficients as $\lambda$.

\subsection{Case II}
The second example of potential is expressed in the following form (T-model attractor),
\begin{equation}\label{24}
U(\psi)=V_{0}\tanh^{2n}(\frac{\psi}{\sqrt{6\alpha}}).
\end{equation}
Like the previous part with respect the equations (\ref{14}) and (\ref{24}), one can obtain,
\begin{equation}\label{25}
\frac{U'}{U}=\frac{2\sqrt{\frac{2}{3}}n Csch[\frac{\sqrt{\frac{2}{3}}\psi}{\sqrt{\alpha}}]}{\sqrt{\alpha}}>C_{1},
\end{equation}
and
\begin{equation}\label{26}
\frac{U''}{U}=-\frac{4n\bigg(-2n+\cosh[\frac{\sqrt{\frac{2}{3}}\psi}{\sqrt{\alpha}}]\bigg)(Csch[\frac{\sqrt{\frac{2}{3}}\psi}{\sqrt{\alpha}}])^{2}}{3\alpha}>-C_{2},
\end{equation}
where $U'$, $U''$ are the first and second derivatives of $\psi$. Also, the tensor-to-scalar ratio and scalar spectral tilt, one can obtain,
\begin{equation}\label{27}
r=\frac{12\alpha}{N^{2}+\frac{N}{2(1-\lambda)}\times\mathcal{M}+3\alpha/4},
\end{equation}
and
\begin{equation}\label{28}
n_{s}=\frac{1-\frac{2}{N}-\frac{3\alpha}{4N^{2}}+\frac{1}{2(1-\lambda)N}(1-\frac{1}{N})\mathcal{M}}{1+\frac{1}{2(1-\lambda)N} \mathcal{M}+\frac{3\alpha}{4N^{2}}},
\end{equation}
where $\mathcal{M}=\sqrt{3\alpha\bigg(4(1-\lambda)^{2}+3\alpha\bigg)}$. As mentioned in the previous section, we will go through a straightforward calculations process to challenge each of the components of the refined swampland dS conjecture, namely $C_{1}$ and $C_{2}$ in terms of two of the most important cosmological quantities $n_{s}$ and $r$. We will rewrite (\ref{27}) and (\ref{28}) inversely and in terms of functions as $N-n_{s}$ and $N-r$, then like the previous part with respect to equations (\ref{4}), and (\ref{25}-(\ref{28}), the swampland conjectures are rewritten according to two very important cosmological quantities, scalar spectral index and tensor-to-scalar ratio as  $C_{1,2}-n_{s}$ and $C_{1,2}-r$.
Now you can see the changes of these parameters in the following figures, which we will analyze the results. After the above calculations, we specify the appropriate range of these cosmological parameters by plotting some figures.

 \begin{figure}[h!]
 \begin{center}
 \subfigure[]{
 \includegraphics[height=6cm,width=6cm]{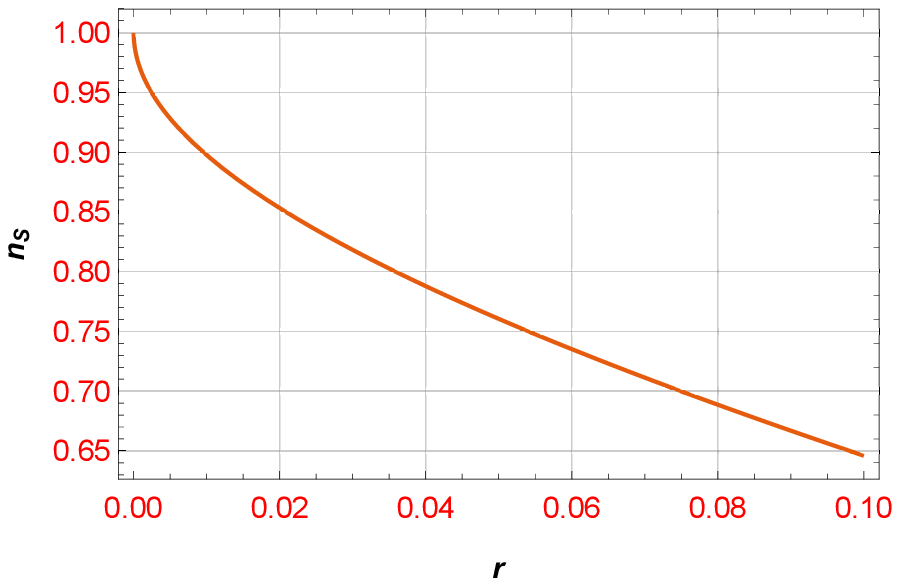}
 \label{14a}}
  \caption{\small{The plot of $n_{s}$ in term of $r$. }}
 \label{4}
 \end{center}
 \end{figure}

 \begin{figure}[h!]
 \begin{center}
 \subfigure[]{
 \includegraphics[height=3.5cm,width=3.5cm]{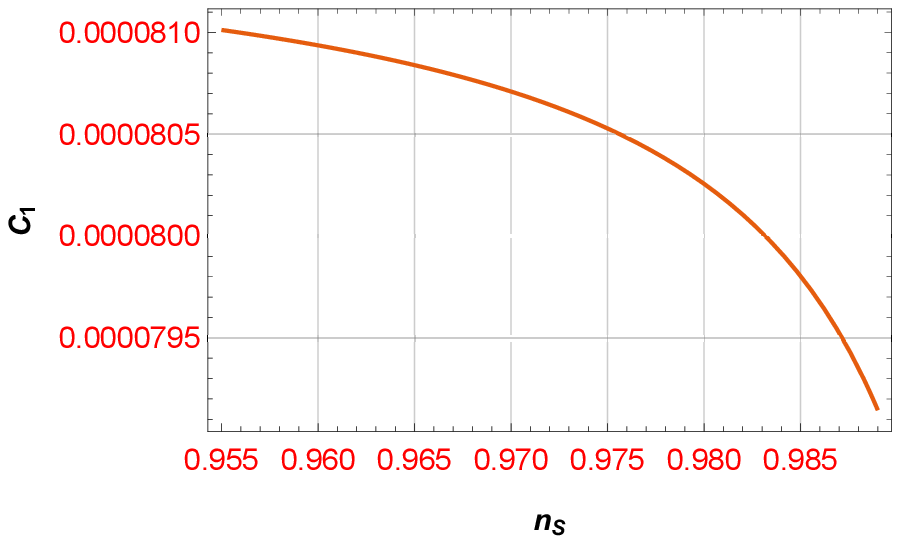}
 \label{5a}}
 \subfigure[]{
 \includegraphics[height=3.5cm,width=3.5cm]{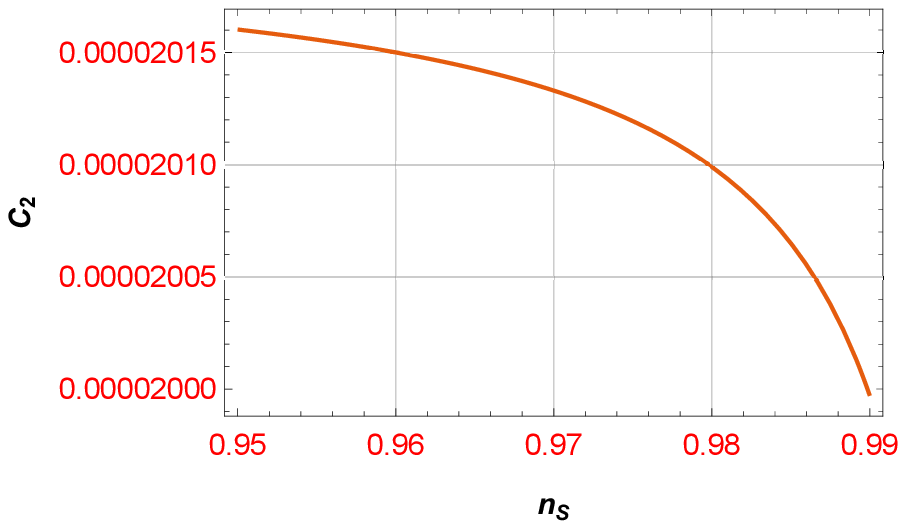}
 \label{5b}}
 \subfigure[]{
 \includegraphics[height=3.5cm,width=3.5cm]{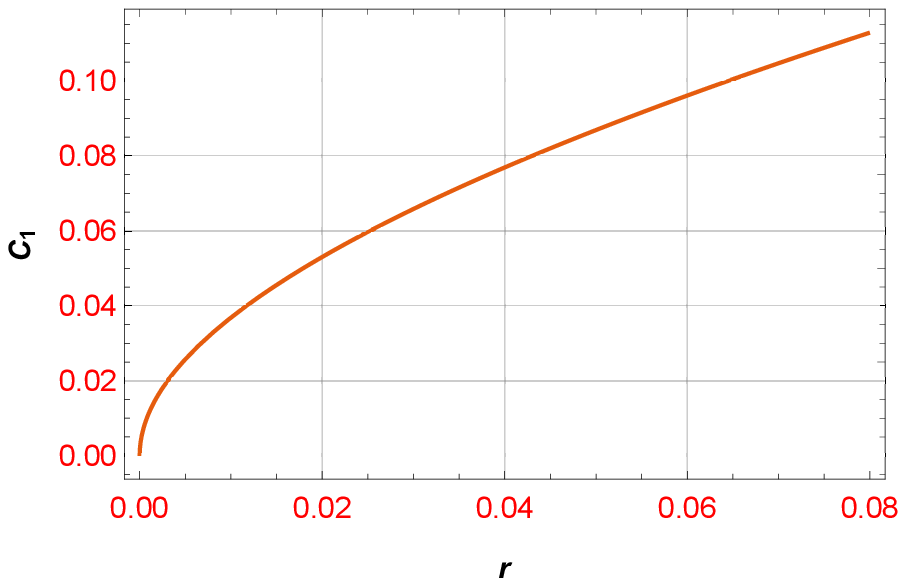}
 \label{5c}}
  \subfigure[]{
 \includegraphics[height=3.5cm,width=3.5cm]{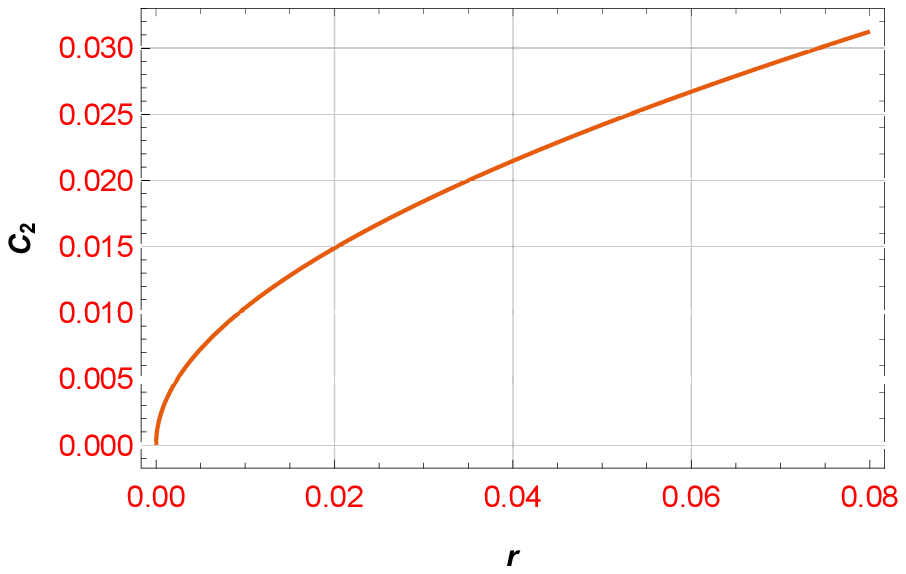}
 \label{5d}}
  \caption{\small{The plot of ($C_{1}$ and $C_{2}$) in term of ($n_{s}$), and ($r$) in Fig. \ref{5a}, \ref{5b} and Fig. \ref{5c}, \ref{5d}, respectively. }}
 \label{5}
 \end{center}
 \end{figure}

 \begin{figure}[h!]
 \begin{center}
 \subfigure[]{
 \includegraphics[height=6cm,width=6cm]{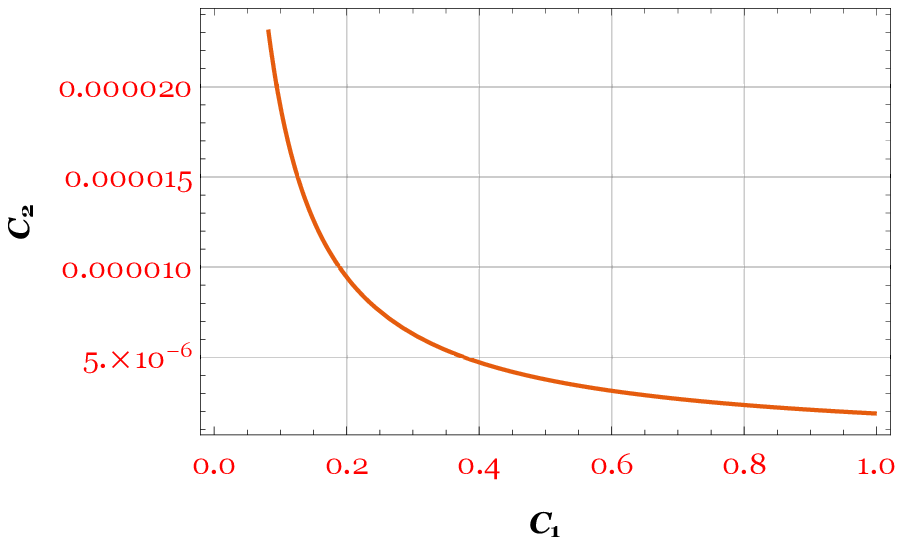}
 \label{6a}}
  \caption{\small{the constraint of ($C_{2}-C_{1}$).}}
 \label{6}
 \end{center}
 \end{figure}

The second model follow the same process as the model (I). But different results are evaluated for each of the parameters. We also describe the results of each of these models in detail. We will discuss the differences and similarities of the mentioned models and introduce the best model compatible with swampland conjectures. We plotted the $r$ in terms of $n_{s}$ in figure \ref{4}. The changes of these two parameters about each other for the mentioned constant parameters are well shown in this figure. The difference can be seen well in comparison with the previous model.
The important point in this model is related to the parameter $\alpha=1-\lambda$, and $n=3$ which expresses good results for the swampland conjectures. different values of $n$, and $\alpha$, also valid the range of swampland conjectures, which is true of all figures. Also, these values viz ($r$) and ($n_{s}$) are within the range of the latest observable data \cite{3}.
Figure \ref{5} shows the changes the swampland conjectures component,i.e., $C_{1}$ and $C_{2}$ in terms of two important cosmological parameters; $n_{s}$, and $r$.  $C_{1}$ and $C_{2}$ in terms of $ n_{s} $ in Fig. \ref{5a}, \ref{5b} and in terms of $r$ in Fig. \ref{5c}, \ref{5d} are plotted regarding  $\alpha=1-\lambda$.
The allowable range of these parameters is well determined.
As shown in the figures, the swampland conjecture components are within the allowable range. As can be clearly seen in the figures, the component $C_{2}$ has smaller values than $C_{1}$, which is, in fact, consistent with the results in the literature and the latest observational data \cite{3}. Like the previous part, we will examine another constraint from the perspective of swampland conjectures and obtain the allowable range, i.e., we investigate the $C_{1}^{2}C_{2}^{2}$ with respect to the latest observable data and the mentioned condition. The allowable range for swampland conjectures of this model is calculated regarding the $\alpha=1-\lambda$.
In figure \ref{6}, we plot the plan ($ C_{2}-C_{1}$) with respect to $\alpha=1-\lambda$.
The allowable range of the swampland conjectures for $\alpha=1-\lambda$, will be $c<0.000012645$. As mentioned above, single-field inflation models with the slow-roll conditions and swampland conjecture have a series of inconsistencies that methods were used to solve this problem in the literature. This paper has resolved these differences and inconsistencies by a known method and small parameter $\lambda$. By introducing several inflation models, we applied each of these concepts to the inflation models. We then obtained each of the cosmological parameters and determined each of these parameters' range by plotting some figures. However, according to the above mentioned, this method can be studied with different theories of gravity and other conditions.

\section{Conclusions}
Recently many researchers have been studied various conditions as the swampland program in cosmological implications. Sometimes these expressed conjectures are modified and lead to fascinating points. Different inflation models, dark energy, and other cosmological implications have also been investigated with swampland conjectures. The swampland conjecture is in contrast to the single-field inflation model with slow-roll conditions. So in this paper, first, we described the special method to solve this problem with the help of a small parameter with respect to a coupling function.  So we introduced some cosmological parameters as the number of e-folds $N$, tensor-to-scalar ratio $r$, and scalar spectral index $n_{s}$ associated with this method, and we explained this issue entirely. Also, we studied a new perspective in inflation models according to the swampland criteria. We introduced some inflation models and also examined the various swampland conditions. We specified the allowable area of cosmological parameters such as ($r$), ($n_{s}$), and reconstructed new restrictions as ($C_{1}-r$), ($C_{1}-n_{s}$), ($C_{2}-r$), and ($C_{2}-n_{s}$).
We determined the allowable range for the swampland component corresponding to the models by plotting some figures and advantaging with the latest observational data \cite{3}. Finally, we expressed the results of these different conditions and compared the result with each other. In studying the models, we identified the permissible areas of swampland conjectures.  These studies found that the contradiction between the swampland conditions and slow-roll single-field inflation can be resolved with these special situations.
The important point is that both models, in addition to compatibility with all the constraints of the swampland dS conjecture, have deeper consistency of these swampland conjecture components with important cosmological parameters such as $r$ and $n_{s}$. The allowable range of swampland conjectures was determined for each model. It was found that swampland conjectures accommodate more volume from the model (I) and are more compatible with this model.
These constraints can also be examined by finding a common area in combining different concepts such as inflation and dark energy, and determining the commonalities of these concepts.
Different inflation models, and the implications of other cosmological studies such as dark energy, etc., can also be evaluated with these new conditions.

\end{document}